\documentclass[a4paper,10pt]{article}
\usepackage{amsmath}
\usepackage{graphicx}

\title{Goldstone bosons in the color-flavor locked phase}
\author{Verena Werth$^\dag$, Michael Buballa$^\dag$, and Micaela Oertel$^\ddag$ \\
{\small \it $^\dag$ Institut f\"ur Kernphysik, Technische Universit\"at Darmstadt, Germany}\\
{\small \it $^\ddag$ Observatoire de Paris-Meudon, France}}
\date{}

\begin{document}
\textwidth=135mm
\textheight=200mm

\maketitle

\begin{abstract}

We study pseudoscalar meson excitations in the color-flavor locked phase within a Nambu--Jona-Lasinio-type model by calculating diquark loops.

\end{abstract}

\section{Introduction}

The ground state of quark matter at highest densities and low temperatures 
is the color-flavor locked phase (CFL) \cite{Alford:1998mk}. In that phase 
the formation of the diquark condensates breaks the original 
$SU(3)_{color} \times SU(3)_L \times SU(3)_R$ symmetry (in the chiral limit) 
down to $SU(3)_{color + V}$. 
Breaking of chiral symmetry leads to the appearance of pseudoscalar
Goldstone bosons.

The properties of the Goldstone bosons have been investigated first
within low-energy effective theory 
approaches \cite{CaGa99,Son:1999cm,Bedaque:2001je}.
These results become exact in the weak-coupling regime at asymptotically 
high densities. 
It has been predicted that kaon and pion condensation may occur
as a consequence of the stress imposed by a non-zero strange
quark mass or non-zero electron chemical potential
\cite{Bedaque:2001je}. More recently, this has been confirmed
within NJL-model approaches
where the meson condensates are realized as non-vanishing
pseudoscalar diquark condensates
\cite{CFLKNJL}. 
In the present contribution we discuss an explicit construction of
CFL Goldstone bosons by calculating diquark loops in an NJL-type model.

\section{Model and formalism}

We use a three-flavor Nambu--Jona-Lasinio model. The Lagrangian density is given by
\begin{equation}
\begin{split}
\mathcal{L}_{e\! f\!f}  = \bar{q} (i \partial\hspace{-1.2ex}\slash - \hat{m}) q
+ H \sum_{A=2,5,7} \sum_{A'=2,5,7} & \Bigl[ \left( \bar{q} i \gamma_5 \tau_A \lambda_{A'} C \bar{q}^T \right) \left( q^T C i \gamma_5 \tau_A \lambda_{A'} q \right) \Bigr.\\
  & \hspace{1ex} + \Bigl.\left( \bar{q} \tau_A \lambda_{A'} C \bar{q}^T \right)\left( q^T C \tau_A \lambda_{A'} q \right)\Bigr]\\
\end{split}
\end{equation}
with the quark field $q$, the diagonal matrix $\hat{m} =
\text{diag}_f(m_u,m_d,m_s)$ of the current quark masses, and the antisymmetric
Gell-Mann matrices in flavor ($\tau_A$) and color space ($\lambda_A$). 
The first term of the Lagrangian is the free part while the second one 
(with the coupling constant $H$) describes scalar and pseudoscalar 
quark-quark interactions. 
In this toy model we do not take into account quark-antiquark interactions.
However, as in the CFL phase baryon number is not conserved, ``mesons''
are essentially diquarks, and it is sufficient to consider quark-quark
interactions only. 

By transforming the Lagrangian into Nambu-Gorkov space, we can extract the diquark interaction vertices
\begin{xalignat}{2}
\Gamma_s^{ll} & = \begin{pmatrix}
0 & 0 \\
i \gamma_5 \tau_{A} \lambda_{A'} & 0 
              \end{pmatrix},  \quad &
\Gamma_s^{ur} & = \begin{pmatrix}
0 & i \gamma_5 \tau_A \lambda_{A'} \\
0 & 0 \\
              \end{pmatrix} \label{scalar_vertices}\\
\Gamma_{ps}^{ll} & = \begin{pmatrix}
0 & 0 \\
\tau_{A} \lambda_{A'} & 0 
              \end{pmatrix},  \quad &
\Gamma_{ps}^{ur} & = \begin{pmatrix}
0 & \tau_A \lambda_{A'} \\
0 & 0 \\
              \end{pmatrix}\; . \label{pseudoscalar_vertices}
\end{xalignat}
For each of these vertices exist 9 combinations of $A$ and $A'$. Altogether we have 18 scalar (eq.(\ref{scalar_vertices})) and 18 pseudoscalar diquark vertices (eq.(\ref{pseudoscalar_vertices})). For the following calculation of the pseudoscalar mesons we only need the pseudoscalar vertices, but of course we require the scalar ones for the CFL gap equations.

In addition we have to provide the inverse propagator in Nambu-Gorkov space
\begin{equation}
S^{-1} = \begin{pmatrix}
p\hspace{-1.0ex}\slash + \hat{\mu} \gamma_0 - \hat{m} & 
        \displaystyle\sum_{A=2,5,7} \Delta_A \gamma_5 \tau_A \lambda_A \\
- \displaystyle\sum_{A=2,5,7} \Delta_A^* \gamma_5 \tau_A \lambda_A &
        p\hspace{-1.0ex}\slash - \hat{\mu} \gamma_0 - \hat{m} 
         \end{pmatrix}
\end{equation}
with the diagonal matrix of the quark chemical potentials $\hat{\mu}$. By
solving the gap equations and neutrality conditions self-consistently, we get
the gap parameters $\Delta_A$ and the color chemical potential $\mu_8$, which
is needed to ensure color neutrality. As, in a first step, we restrict 
ourselves to zero temperature we do not need the color chemical potential
$\mu_3$ and the electric charge chemical potential $\mu_Q$. For the finite
temperature case these have to be included (see, e.g., \cite{Ruster:2005jc}).

Our aim is to solve the Bethe-Salpeter equation
\begin{equation}
 \mathcal{T}=\mathcal{V}+\mathcal{VJT}.
\end{equation}
The main ingredient is the polarization function $\mathcal{J}$ which is given 
by
\begin{equation}
  -i \mathcal{J}_{jk}(q)  = - \int \frac{d^4 p}{(2 \pi)^4} \frac{1}{2} \textnormal{Tr}
        \lbrack \Gamma_j^{\dagger} i S(p+q) \Gamma_k i S(p) \rbrack\; .
\end{equation}
Here $\Gamma_j$ and $\Gamma_k$ are the pseudoscalar diquark vertices given in eq.~(\ref{pseudoscalar_vertices}). To solve this expression, we apply the Matsubara formalism and restrict ourselves to the case $\vec{q}=0$. This leeds to
 \begin{equation} 
  -i \mathcal{J}_{jk}(i\omega_m,\vec 0) = \frac{i}{2} T \sum_n \int \frac{d^3 p}{(2 \pi)^3}    \frac{1}{2} \textnormal{Tr} \lbrack \Gamma_j^{\dagger} i S(i \omega_n + i \omega_m,\vec{p})
        \Gamma_k i S(i \omega_n, \vec{p}) \rbrack \; ,
 \end{equation}
with the fermionic Matsubara frequencies $\omega_n=(2\pi+1)nT$.

Only a few vertex combinations lead to non-vanishing $\mathcal{J}_{jk}$. 
This enables us to choose a basis in which $\mathcal{J}$ is block-diagonal. 
In this basis $\mathcal{T}$ is also block-diagonal, which makes the 
calculation much simpler. We are left with six $2\times 2$ blocks and one 
$6\times 6$ block. Each of the small blocks can be identified with one of the flavored mesons ($\pi^+$, $\pi^-$, $K^+$, $K^-$, $K^0$, 
and $\bar{K}^0$). The unflavored modes ($\pi^0$, $\eta$, and $\eta'$) 
mix with each other and are contained in the $6\times 6$ block.
We leave the study of this block for future work.

\section{Results}

In the following we discuss our results for $\mu = 500$\,MeV. The
model parameters are given by a 3-momentum
cutoff $\Lambda = 602.3$\,MeV and $H\Lambda^2 = 1.37625$, corresponding to 
a CFL gap of approximately $77$\,MeV.

\begin{figure}[hbt]
\begin{center}
 \includegraphics[width=0.49\linewidth]{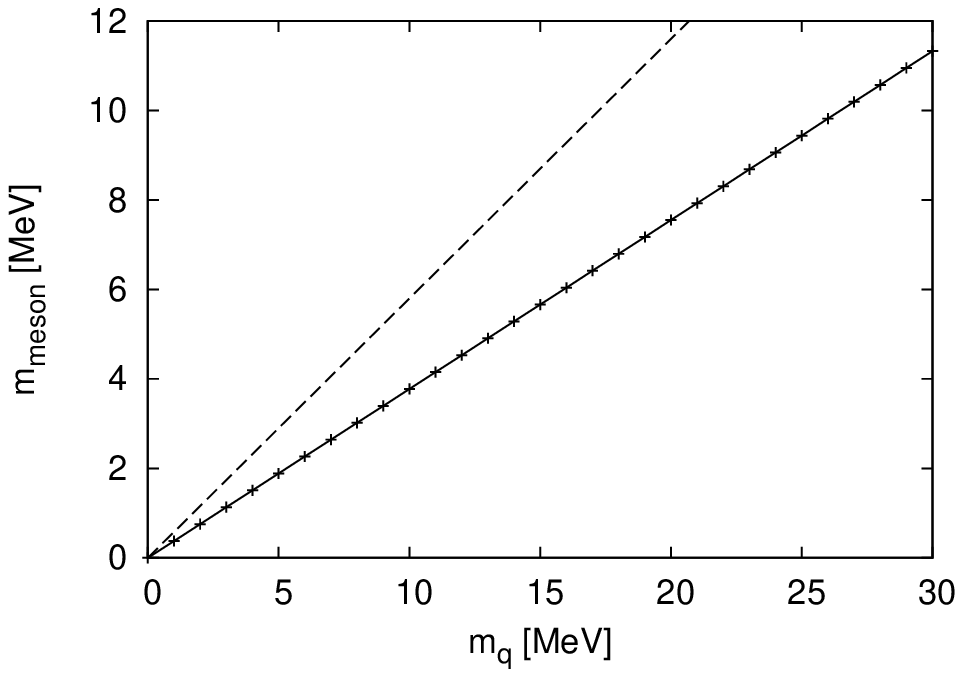}
 \includegraphics[width=0.49\linewidth]{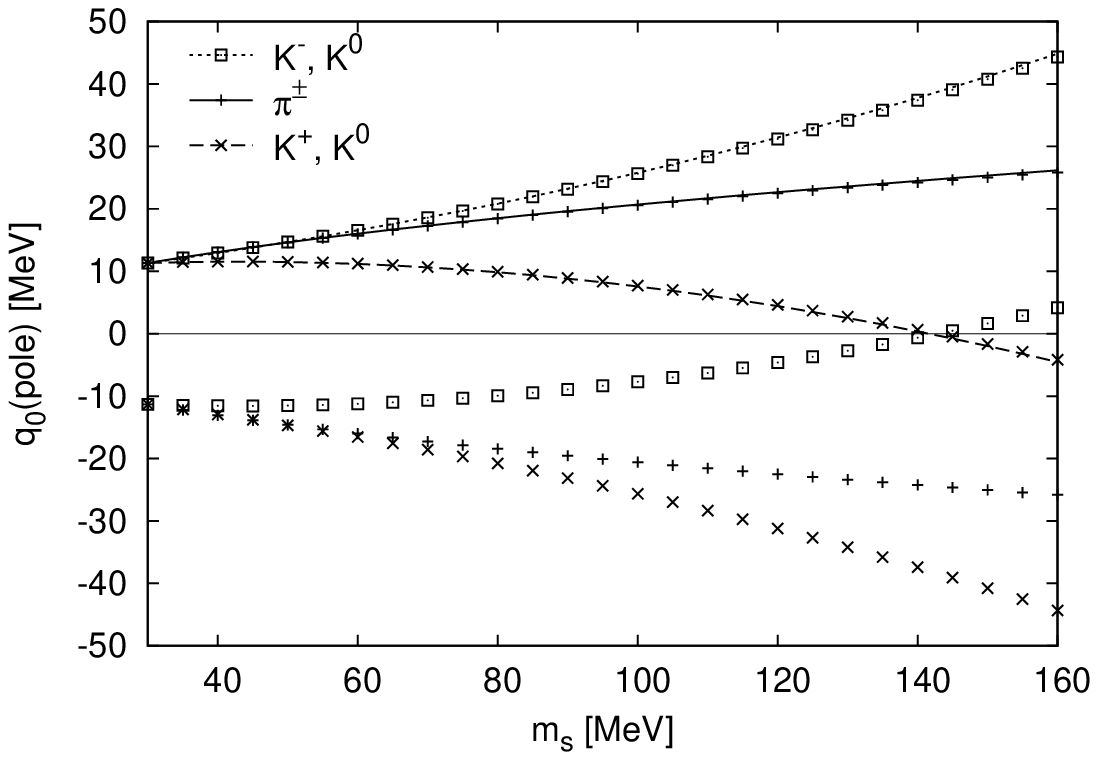}
 \caption{Meson properties for $T=0$ and $\mu=500$\,MeV. 
\textit{Left}: Meson mass as a function of a common quark mass, 
solid line: our calculation, dashed line: result from EFT. 
\textit{Right}: Pole positions as functions of the strange quark mass for 
$m_u=m_d=30$\,MeV, points: our calculation, lines: fit with EFT.}
 \label{meson_masses}
\end{center}
\end{figure}

We begin with the case of equal masses for up, down, and strange quarks. 
In this case, all considered meson masses are degenerate and
given by the $q_0$-values at which the poles in $\mathcal{T}$ occur. 
The results are displayed in the left panel of Fig.~\ref{meson_masses}.
We find a linear dependence of the meson mass on the quark mass $m_q$.
This was also derived in \cite{Son:1999cm} in a low-energy effective field 
theory (EFT) approach. There the authors give an expression for the slope 
$a$ of the curve:
\begin{equation}
 a_{\textit{\scriptsize{EFT}}}=\sqrt{\frac{8A}{f_{\pi}^2}}
 \quad \textnormal{with} \quad
   A=\frac{3\Delta^2}{4\pi^2}\quad \textnormal{and} \quad
   f_{\pi}^2=\frac{21 - 8 \ln 2}{18} \frac{\mu^2}{2 \pi^2}\; .
   \label{A_and_fpi}
\end{equation}
On the left-hand side of Fig.~\ref{meson_masses} this prediction is plotted as 
a dashed line. The slope of our calculation is approximately 35\% smaller than 
the one obtained in EFT. 
We have repeated this comparison for smaller values of $\Delta$,
i.e., lowering the interaction strength $H$. 
The deviation then gets reduced.
This is expected because eq.~(\ref{A_and_fpi}) was derived for weak 
coupling.

As a next step we keep $m_u$ and $m_d$ constant at 30\,MeV and increase $m_s$. 
The previously degenerate solutions for the poles of $\mathcal{T}$ then
split into three branches, which are the points plotted on the 
right-hand side of Fig.~\ref{meson_masses} as a function of the strange quark 
mass (points). 
Since we still have chosen equal masses for up and down quarks, 
$\pi^-$ and $\pi^+$, $K^-$
and $K^0$, and $K^+$ and $\bar{K}^0$ remain degenerate.

Again we compare our results with those derived from EFT \cite{Bedaque:2001je}:
\begin{align}
    q_{\pi^{\pm}}(pole) & = \mp \frac{m_d^2 - m_u^2}{2 \mu} 
    + \sqrt{\frac{4A}{f_{\pi}^2} m_s (m_u + m_d)}
\nonumber\displaybreak[0]\\
    q_{K^{\pm}}(pole) & = \mp \frac{m_s^2 - m_u^2}{2 \mu}
      + \sqrt{\frac{4A}{f_{\pi}^2} m_d (m_u +
        m_s)}
\nonumber\displaybreak[0]\\
    q_{K^0,\bar{K}^0}(pole) & = \mp \frac{m_s^2 - m_d^2}{2 \mu}
      + \sqrt{\frac{4A}{f_{\pi}^2} m_u (m_d + m_s)}\; .
\label{mesonseft}
\end{align}
The first term on the right hand side can be interpreted as 
(the negative of) an effective chemical potential of the corresponding meson, 
the second term is the mass of the meson. 
For the comparison with our results we replace the terms
$\sqrt{4A/f_{\pi}^2}$ in eq.~(\ref{mesonseft}) by
$a/\sqrt{2}$, where $a$ is the slope of our calculation shown in the left 
panel of Fig.~\ref{meson_masses}. The results are displayed as lines 
in the right panel. They agree almost perfectly with our calculations. 

For the chosen parameters, $q_0(pole)$ becomes zero for $K^+$ and $K^0$
at $m_s \approx 142\,\text{MeV}$. At this point kaon condensation sets in. 
For higher strange quark masses the CFL phase is no longer the correct ground 
state and the shown results have no physical meaning. To study mesonic
excitations in this regime, one should first calculate the CFL+K ground 
state \cite{CFLKNJL}.

\section{Summary \& Outlook}

We have described mesons with diquark loops. Our results are in good 
qualitative and -- for weaker couplings -- even in good quantitative
agreement with high-density effective theories.

Much work remains to be done: First, we should also study 
$\pi^0$, $\eta$, and $\eta'$.
Second, we should include more interactions,
e.g., quark-antiquark interactions, which are responsible for
dynamical mass generation, or
instanton induced interactions, which are expected to increase the
meson masses considerably. 
Finally, we intend to calculate the mesonic excitations in the 
CFL+meson ground state.

\vspace{-1.5ex}
\subsubsection*{Acknowledgments}
\vspace{-1.2ex}
V.~W.~thanks the program HISS Dubna funded by the Helmholtz Association jointly with its institutes DESY and GSI for her participation at the Summer School. M.~B.~acknowledges support by the Heisenberg-Landau program.

\end{document}